\documentclass[conference, letter]{IEEEtran}
\IEEEoverridecommandlockouts
\usepackage{cite}
\usepackage{amsmath,amssymb,amsfonts}
\usepackage{algorithmic}
\usepackage{graphicx}
\usepackage{balance}
\usepackage{textcomp}
\usepackage[table]{xcolor}
\usepackage{float} 
\usepackage[USenglish]{babel}
\def\BibTeX{{\rm B\kern-.05em{\sc i\kern-.025em b}\kern-.08em
    T\kern-.1667em\lower.7ex\hbox{E}\kern-.125emX}}
\begin{document}

\title{MCQUIC - A Multicast Extension for QUIC\\
}

\author{\IEEEauthorblockN{Max Franke}
\IEEEauthorblockA{
\textit{TU Berlin}\\
Berlin, Germany\\
mfranke@inet.tu-berlin.de}
\and
\IEEEauthorblockN{Jake Holland}
\IEEEauthorblockA{
\textit{Akamai}\\
Pasadena, USA\\
jholland@akamai.com}
\and
\IEEEauthorblockN{Stefan Schmid}
\IEEEauthorblockA{
\textit{TU Berlin \& Fraunhofer SIT}\\
Berlin, Germany\\
stefan.schmid@inet.tu-berlin.de}
\thanks{This work has been accepted for publication at NCA 2024 under DOI 10.1109/NCA61908.2024.00037. The authors acknowledge the financial support by the Federal Ministry of Education and Research of Germany in the programme of ``Souver\"an. Digital. Vernetzt.'' Joint project 6G-RIC, project identification number: 16KISK030.}
}


\maketitle

\begin{abstract}
Multicast presents an attractive and efficient solution for streaming large-scale live media events, such as the Olympics, over the Internet.  However, existing solutions face several drawbacks, particularly concerning security and privacy, which hinder their implementation in web browsers.

In this paper, we introduce MCQUIC, a multicast extension to the QUIC transport protocol which addresses many of these challenges. Utilizing existing network layer multicast mechanisms, MCQUIC augments multicast delivery by providing encryption and integrity verification of packets distributed over multicast, along with automatic unicast fallback. Moreover, it operates transparently to applications and can easily be utilized by enabling a simple option in QUIC. We analyze and compare it to other delivery methods for live video, both theoretically and empirically. We are currently in the process of specifying MCQUIC at the IETF.   
\end{abstract}

\section{Introduction}
Major live events such as world cups, the Superbowl, or the Olympics attract audiences of hundreds of millions of viewers. While traditionally consumed on TV, an increasing number of viewers now follow such events on the Internet. Streaming these peak-traffic events via unicast is challenging and costly for both Internet service providers (ISPs) as well as content delivery networks (CDNs)\cite{marcon2011local}.
 Using multicast instead would reduce load on many parts of the network infrastructure by removing the need to send the same exact data multiple times to different receivers. 

Two concrete examples where unicast delivery is reaching its limit are live streaming and game downloads. As of July 2024, the highest traffic peak that Akamai, one of the worldwide largest CDN providers, recorded across its network was 250 terabit per second \cite{akamai_2022}. A 4K livestream has a bitrate of around 40 Mbps~\cite{4kBitrate}. This means that 6.25 million concurrent viewers are enough to fully utilize Akamai's available capacity. This is less than 2\% of the average viewership of e.g. the EURO finals \cite{uefa}. This is not even considering an increase in bitrate due to HDR or 8K. Similarly, new game releases can cause huge bursts of bandwidth demand. One such release is GTA V. As the fastest selling game of all time, it sold over 11 million copies on the first day \cite{GTAV}. With a file size close to 100GB~\cite{GTASize}, it would take all of Akamai's capacity almost 10 hours to handle the downloads on its launch day. Since its release, file size has also continuously grown, now reaching up to 231GB for certain games~\cite{Gamesizes}. Both of these cases illustrate that already today unicast delivery is becoming unsustainable, while traffic demand is certain to further increase. This might soon lead to an inflection point where increases in link capacities can no longer keep up with demand.  CDN providers especially are aiming to remove customers that cause this, usually unprofitable, peak-traffic\cite{Rayburn_2024}.

While multicast is widely used for intra-domain use cases, such as mDNS \cite{rfc6762}, its inter-domain applications are very limited. Even though there exists a multicast backbone, called the MBONE, it is mostly comprised of educational networks such as Internet2 and GÉANT, while carrier and ISP networks usually disable generic native inter-domain multicast. This was and is largely due to a lack of legitimate applications that use multicast, a lack of privacy and security \cite{hardjono2000ip} when compared to unicast, and the large amounts of overhead required to create and maintain any-source multicast (ASM) trees \cite{diot2000deployment}.

\begin{figure}[t]
  \centering
  \includegraphics[width=0.5\linewidth]{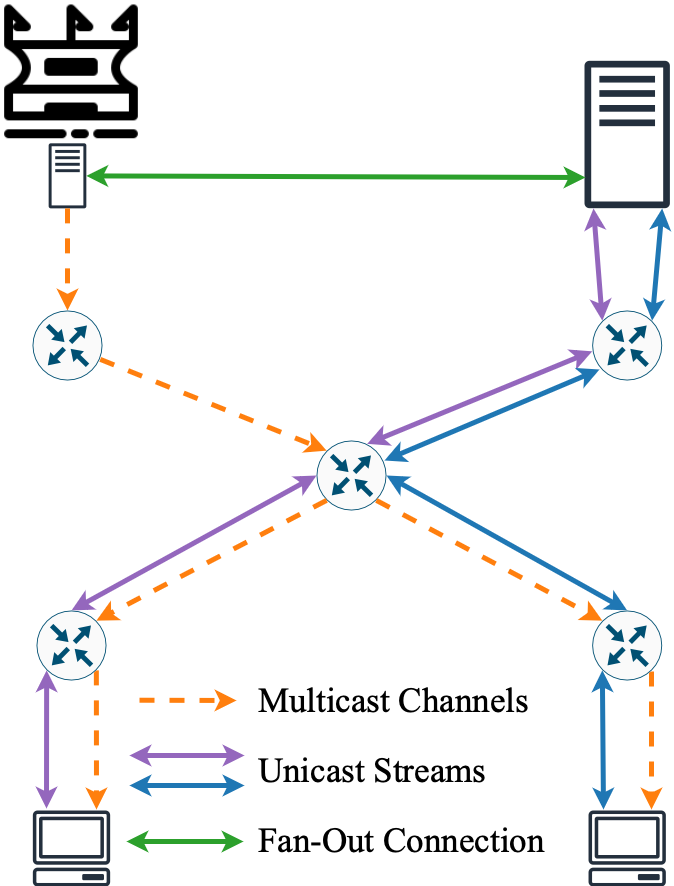}
  \caption{MCQUIC combines QUIC unicast connections with multicast channels to allow for scalable and robust delivery of mass media content. Schematic overview of a multicast source, a Fan-Out server and two clients using MCQUIC. The orange connection represents multicast, used to carry big data packets such as video frames, while the connections in the purple and blue represent unicast connections, used only for control. The multicast source does not have to be co-located with the server handling the unicast connections. Instead, a fan-out connection, displayed in green, can be used to allow for load balancing.}
   \label{fig-schema}
\end{figure}
However, recent developments have eased these issues. For one, ASM has been officially deprecated \cite{rfc8815} for inter-domain use cases, removing much of the complexity related to finding multicast sources. On the other hand, the traditional protocols used to maintain multicast trees, most notably PIM \cite{rfc7761}, are partly being replaced by BIER \cite{rfc8279} which provides a way to route multicast packets without the need to maintain trees and thus state on each router. Furthermore, automatic multicast tunneling (AMT) \cite{rfc7450} provides a way to bridge networks that do not support native multicast. As such, the time might have come to make another attempt to deploy multicast content delivery to the web. 



In this work, we introduce MCQUIC which is an extension to the QUIC transport protocol, which was initially proposed by Google in 2012 and standardized by the IETF in 2021. MCQUIC leverages QUICs packet protection and multiple streams to combine unicast and multicast. 
QUIC itself is already widely implemented in most popular browsers, allowing for a feasible way to deploy MCQUIC. By combining QUIC and multicast, we can leverage the security benefits inherent in unicast communication, as well as the use as a fallback channel, with the scalability benefits of multicast. 

Our extension allows clients to use encrypted multicast that is protected against injection of packets by third parties. This enables trust in packets received via multicast on public networks. It only requires minimal changes to application layer implementations and offers automatic fallback to unicast QUIC in case the network does not support multicast. 

It should be noted that the extension uses already existing mechanisms and technologies to realize multicast on the network layer and is exclusively focused on the transport layer. 
We will compare our appraoch with other existing transports, namely HTTPS for DASH/HLS and native multicast, both theoretically as well as experimentally, showing that MCQUIC finds a balance between the security and privacy benefits of the first and the scalability benefits of the second. We show that up to 4 times as many clients can be served when using our approach as compared to strict unicast delivery. 
\section{Motivation and design overview}
\label{design}
In this section we will give an example to motivate why MCQUIC is a useful addition to QUIC before giving some high level information on its design principles. It should be noted that this paper focuses on use cases and analyzing the performance benefits and trade-offs of MCQUIC. While we will highlight some of the major design decisions later in section \ref{detail_design}, the full details are available in the proposed specification \cite{jholland-quic-multicast-04}. 
\subsection{Motivating example}
One of the main potential use cases for MCQUIC is the streaming of live media in browsers. As mentioned before, more and more people shift their content consumption from traditional (cable) TV to live streaming. As such, the number of people that are going to watch live sports digitally in the US alone is expected to increase from 57.5 million in 2021 to over 90 million by 2025 \cite{liveSports}. This often either occurs in browsers directly or in browser-based apps on smart TVs or devices like Chromecast or AppleTV.

Figure 1 illustrates a deployment scenario showcasing the application of MCQUIC in the live broadcast of a sports event.
The Multicast Source (MCS) is positioned in the top left corner, in this case representing the broadcast studio of a football stadium.

In the top right, a Fan-Out Server (FOS) is shown which will manage the unicast aspect of MCQUIC. While our example features a single server, real-world deployments may incorporate multiple servers or an entire CDN.

The MCS provides information to the FOS, including checksums for multicast packets and cryptographic data necessary for decryption through a Fan-Out connection. Additionally, the MCS ingests video segments of the live broadcast into the network by sending them to a specific multicast channel. There might be multiple channels, e.g. one for 1080p and one for 4K.

A user now tries to view the game by accessing the live stream page via their browser. Initially, the user's browser establishes a connection to the FOS through QUIC unicast. During this connection setup, the client communicates its multicast-related preferences to the FOS. Based on these, the FOS identifies the most suitable multicast channel for the user. 

\begin{figure}[t]
  \centering
  \includegraphics[width=1\linewidth]{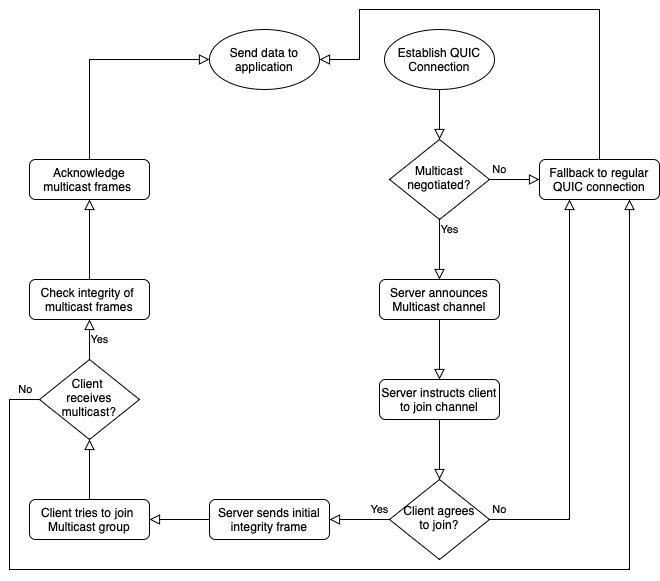}
  \caption{Simplified flow diagram of an MCQUIC connection. }
   \label{fig-flow}
\end{figure}

Subsequently, the FOS directs the client to join the selected channel, by subscribing to the multicast group to which the MCS is sending the live stream. Furthermore, the FOS forwards information received from the MCS securely to the client via the unicast QUIC connection. Upon successful joining of a multicast channel, the client decrypts packets and verifies their integrity by calculating a checksum over them and comparing it to the ones provided by the FOS. It signals its receipt of packets by acknowledging them like regular QUIC packets to the FOS. In the event of no acknowledgments within a specified time-out, the FOS may infer that the client is incapable of receiving multicast traffic. In such cases, the FOS seamlessly transitions to transmitting subsequent video segments directly via unicast, ensuring uninterrupted delivery with immediate fallback capability. Independent of the way the data was received, the browser can then start to show the video to the user. This is shown in the simplified flow diagram in figure \ref{fig-flow}.

By only requiring modifications inside of QUIC, this approach allows for easy adoption by applications as they have to make little to no changes inside of their client-side implementations. Additionally, thanks to the automatic fallback to unicast, there is little risk in the use of MCQUIC. If multicast is available, it will enable great scalability improvements while performing the same as any other unicast connection if it is not. 

\subsection{Design overview}

The primary design enhancement introduced by MCQUIC is the capability for clients to join one or more multicast QUIC channels in addition to the conventional unicast QUIC connection. These channels consist of unidirectional data streams from the server to the client. 

For simplicity's sake, moving forward, we will assume that the FOS and MCS are co-located and will simply refer to them collectively as `the server`. Unless stated otherwise, control frames are sent via unicast. 
The QUIC connection is established following the standard procedure. It undergoes the TLS handshake and certificate chain verification to authenticate the server to the client and exchange the necessary keying material for encrypting the connection.\footnote{The exact handshake process, like many other aspects of QUIC, falls outside the scope of this document but can be referenced in RFC 9000 \cite{rfc9000} and RFC 9001 \cite{rfc9001} respectively.}

During the handshake, the server and client exchange transport parameters, which define the capabilities and constraints of each peer.
MCQUIC introduces two new transport parameters. These are used to negotiate support for multicast. Additionally, the client utilizes its transport parameter to convey its own requirements and limitations concerning multicast to the server.

A server can instruct a client to join a specific channel carrying content that is desirable for the client to receive. However, whether a channel is joined ultimately rests with the client; it reserves the right to refuse joining a channel for various reasons, such as concerns about violating its own bandwidth limitations. If a client opts to join a channel, it notifies the server of this decision by transmitting a relevant frame over the unicast connection.

A client would then proceed to join the multicast channel utilizing standard existing technologies, such as IGMP\cite{rfc3376} and MLD\cite{rfc3810}. Upon successful join, packets will start to arrive at the client, which will acknowledge their reception to the FOS over unicast.

All packets transmitted via multicast are encrypted. Their integrity is also guaranteed by a corresponding checksum that is securely transferred. This mechanism enables the client to verify the authenticity of each packet, ensuring it originated from the original source and was not modified or injected by a third party. The computational overhead incurred during packet verification depends on the choice of hash function employed, we will give some analysis on this in Section \ref{Eval}.

Following the verification of a multicast packet's integrity, the client processes it the same way it would a regular unicast packet. Consequently, to the application layer, it is transparent if packets have been received via multicast or unicast.
\subsection{MCQUIC capabilities and properties}
To summarize, MCQUIC offers the following capabilities and properties:
\begin{itemize}
    \item Reception and processing of packets delivered via multicast within a QUIC connection.
    \item Transparency for applications utilizing QUIC as their underlying transport as to whether a packet was delivered via unicast or multicast.
    \item Automatic fallback to unicast transmission in instances where multicast reception is not possible.
    \item Packet encryption and integrity assurances for packets transmitted via multicast.
    \item The option for both reliable (in the form of STREAM frames) and unreliable (in the form of DATAGRAM frames) delivery of multicast packets.
    \item Clear separation of concerns between the multicast source and servers managing unicast connections, allowing load balancing and CDN usage.
    \item A server-driven but client-executed design, allowing clients to control the amount of data they receive based on their capabilities. 
\end{itemize}
\section{Challenges solved}
\label{challenges}
As stated previously, traditional multicast deployments face challenges concerning security, privacy, and the lack of support for native multicast in many (ISP) networks. In this section, we will provide a more comprehensive explanation of these issues and elaborate on how our approach solves them.

\subsection{Security and privacy}
MCQUIC allows decryption keys to be distributed via the unicast connection, thereby ensuring that packets transmitted over multicast are protected by the same mechanisms used for regular unicast QUIC packets. As such, third parties are unable to gain any additional information on the content carried on a specific multicast channel.

However, since identical keying material is used by all receivers, an attacker could access the content of the channel if they obtain the keys from any legitimately subscribed receiver. This highlights the importance of regularly updating the keying material to keep forward secrecy. The responsibility falls on the server to implement access controls that ensure that only legitimate clients receive the keying material.

By using the same encryption mechanisms used by other QUIC implementations, we can also utilize any pre-existing hardware offloading capabilities for cryptographic functions.

\subsection{Prevention of packet injection}
Multicast senders typically lack knowledge of the number, let alone the identity, of receivers. As a result, there exists no secured two-way communication between them. Receivers therefore can not verify whether a packet received on a multicast channel was indeed sent by the original sender or by a third party spoofing its IP address.

In MCQUIC, we address this issue with the addition of integrity frames which are transmitted, at least initially, over a secure unicast QUIC connection. The checksums used to guarantee integrity utilize standard cryptographic hashes that are resilient against collision attacks. Accordingly, receivers are capable of verifying each packet received over multicast against its corresponding checksum. Upon a successful match, receivers can guarantee that the packet was indeed sent by the intended sender and was not injected by a third party. While the initial checksums need to be transmitted via unicast, subsequent ones can be send on the multicast channel by utilizing a Merkle tree.

\subsection{Fallback to unicast}

Currently, the majority of (ISP) networks do not support generic native multicast. Consequently, applications that want to utilize multicast must implement a unicast fallback mechanism at the application layer.

With MCQUIC, applications only need to indicate whether they wish to allow multicast reception or not. From there on, the QUIC implementation handles the fallback process if needed. If multicast is unsupported by the network, MCQUIC seamlessly transitions to unicast and functions like a standard QUIC connection. This fallback mechanism, depending on the QUIC implementation, remains transparent to the application layer.
\section{Design decisions}
\label{detail_design}
In this section we will give some more details on some of the designs of MCQUIC. The three specific decisions we will elaborate on are the need for multicast specific transport parameters, how multicast channels fit into QUIC connection and how we envision flow and congestion control to be handled. 
\subsection{Client limits}
During connection establishment, both client and server can include an additional multicast transport parameter that indicates to the peer that multicast is supported. The client's parameter includes additional details, such as a list of supported hash and encryption algorithms and whether or not IPv4, IPv6, or both are supported for multicast reception. These parameters can be later updated by the client by sending an LIMITS frame to the server that contains the same parameters. 

There are several reasons as to why we need separate transport parameters instead of using the ones already negotiated for unicast QUIC. Even though there might be support for both IPv4 and IPv6 via unicast, the client might be aware that due to a lack of MLD support, only IPv4 multicast is supported. In this case, disallowing IPv6 multicast from the get go saves time when the server would otherwise instruct the client to join an IPv6 multicast channel only to find out no reception is possible. 

The list of hash and encryption algorithms might also differ from unicast QUIC as it can be assumed that only less sensitive data will be carried over multicast. This means that there might be an option or desire to use weaker algorithms, such as SHA-128, to reduce the size of the checksums carried in integrity frames. While it is true that SHA-128 is no longer considered to be entirely collision resistant \cite{leurent2020sha}, it might be a worthwhile trade off to still use it in certain cases where it is highly unlikely that an attacker can create a collision in time or where a collision would not pertain to any critical or sensitive data. In either case, it is always recommended to use collision resistant algorithms to prevent any such risks from occurring at all. 

Also included in the client side transport parameter are the limits the client places on all multicast channels combined. These include the maximum number of announced or simultaneously joined channels, but most importantly the maximum aggregated rate across all channels. As the data carried over multicast is often of a steady and pre-determined rate, e.g. in video streams the bitrate is known beforehand, the server can reliably take restrictions of clients into consideration when assigning which channels to join. It might determine that the maximum allowed rate is not high enough to receive a 4K stream, which gets carried on one channel, so instead instructs the client to join a different channel that carries the stream in HD. More sophisticated models could make use of layered video \cite{huo2015tutorial} or similar to provide each client with an optimal utilization of its available bandwidth. 
\subsection{Multicast channels}
In QUIC, clients and servers assign (multiple) connection IDs to each connection. These are included in each sent packet. This mechanism is used to enable mobility by making a connection identifiable even if the 5-tuple changes. For example, any packet sent from server to client contains a client selected ID so the client can associate it with a connection. In multicast QUIC channels, we use the connection ID field in packets and replace it with a channel ID that has the same format. This allows clients to associate each packet with a multicast channel, instead of a connection. Connection and channel IDs share the same space. For clients to learn that an ID is a channel ID and not a connection ID, the server sends channel announce frames that contain, among other parameters, channel IDs of available multicast channels.

These channels are abstractions that represent a source specific multicast channel along with additional parameters such as the encryption algorithm used for header and packet protection. The secret for the header algorithm is also included and remains static during the lifetime of a channel while the secret for packet protection is sent separately in a KEY frame. Channels exist independent of any single unicast connection and may be joined by many clients. Their lifetime can range from a few seconds to days or even weeks. Channels share the stream ID space with each other and the unicast connection. This allows for the retransmission of lost packets over a different channel or unicast. Contrarily, each channel has its own packet number space to reduce the coordination required between channels.
In addition to the ANNOUNCE and KEY frames, the server will also send INTEGRITY frames that include hashes for each packet sent over the multicast channel. These are used to authenticate packets and as the hash algorithms used are secure against collision attacks, this mechanism protects against the injection of packets into the multicast channel by third parties.

A server will then at some point send a JOIN frame that will indicate to the client that it is should now join the specified channel. If it decides to do so, it will achieve this by e.g. sending an IGMP or MLD report which triggers the joining of the SSM channel. If the network supports multicast and packets start to arrive, the client will send a STATE frame to the server indicating this. As the entire multicast process is server driven, the server might decide that if a specific timeout passes without receiving a STATE frame, that the clients network does not support multicast. In that case, it can fall back to unicast and send the data packets over the regular QUIC connection \cite{gruessing-moq-requirements-03}. 

Any packets arriving on a multicast channel get acknowledged by the client over the unicast connection. This is necessary as the channels are strictly unidirectional from server to client. This also means that only specific frames, such as STREAM or DATAGRAM frames, can be sent on them. To reduce load and improve scalability, acknowledgements can be bundled and don't need to happen immediately. Lost packets may be retransmitted. This can occur over both unicast or multicast, depending on what the server deems more efficient. 
\subsection{Congestion and flow control}
As there is no regular congestion or flow control mechanism on channels, the client specifies a maximum allowed aggregated rate across all channels. Depending on this rate, the server decides on which channels are suitable for the client to join to make sure that all joined channels combined stay under the clients limits. Each channel ANNOUNCE frame also includes the maximum rate of data that is expected to be sent on that channel. If the client thinks that joining a channel would violate its limits, it can refuse the join. A client can also update its allowed maximum aggregated rate, along with other parameters, during the lifetime of a connection by sending a LIMITS frame. Both server or client may decide that the client should leave a joined channel for a variety of reasons. These include, but are not limited to, high persistent loss on the channel, the end of the channels lifetime, or excessive amounts of spurious traffic (that may be injected by an attacker). 
\subsubsection{Flow control}
The usual mechanisms used for flow control in QUIC are too strict for use in multicast channels without modification. The multicast extension gives the client a new responsibility to be able to robustly handle multicast packets that would exceed its MAX\_DATA without aborting the connection, either by increasing its MAX\_DATA as needed to keep up with received multicast packets or by dropping the packet and leaving the channel (resulting in unicast fallback), for clients that cannot do so. As there are potentially many clients joined on a single channel, the server can not adjust its sending of data by limits imposed by any individual one. Additionally, the new transport parameter is used by clients to set their limits and change them by sending new Limit frames. The server has to make sure that the client stays within those limits by having it either join or leave channels appropriately. 
\subsubsection{Congestion control}
The server is aware of any multicast packet loss experienced by the client in the form of missing MC\_ACK frames. The server should make sure that clients leave channels that experience heavy sustained loss. If several clients that are subscribed to the same channels experience loss of the same packets, this might indicate an underlying issue further upstream. Contrarily, if a client experiences loss on several channels at once it might indicate an issue on its own end. As such, it should react by lowering its max rate parameter. 
\section{Analysis}
\label{Eval}

In this section we will analyze and compare MCQUIC to two other potential delivery methods for live video. The two main focus areas that can cause issues with scalability are bandwidth usage and processing/memory overhead. We will calculate how much bandwidth the different methods of integrity verification require when used within MCQUIC. Additionally, we will show the results of an experiment that evaluates how many clients can be served in parallel by the different delivery methods. 
\subsection{Comparison to other transports}
\begin{figure}[h]
  \centering
  \includegraphics[width=0.8\linewidth]{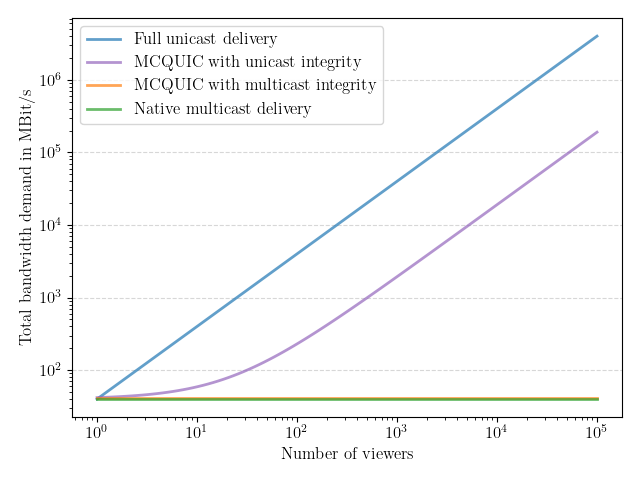}
  \caption{Bandwidth usage for total number of viewers of a 4K live video. Both native multicast and MCQUIC with multicast integrity remain at a constant bandwidth while MCQUIC with unicast integrity scales around 1.5 magnitudes better than full unicast video delivery.}
   \label{fig-subs}
\end{figure}

We will go back to the use case from Section \ref{design}, the live streaming of a sports game, and compare MCQUIC to both native multicast over UDP as well as HTTPS over TCP, which is commonly used today for live streaming with DASH or HLS. 
Table \ref{tab:comp} shows the characteristics of these 3 methods, namely in regards to four metrics we considered.
\begin{table}[h]
\centering
\begin{tabular}{|l|c|c|c|c|}
\hline
\rowcolor[HTML]{E0E0E0} 
\textbf{Transport} & \textbf{Security} & \textbf{Privacy} & \textbf{Deployability} & \textbf{Scalability} \\ \hline
Native multicast & \cellcolor{red!25} Low  & \cellcolor{red!25} Low & \cellcolor{red!25} Low & \cellcolor{green!25} High    \\ \hline
HTTPS    & \cellcolor{green!25} High  & \cellcolor{green!25} High & \cellcolor{green!25} High & \cellcolor{red!25} Low    \\ \hline
MCQUIC   & \cellcolor{green!25} High  & \cellcolor{yellow!25} Medium & \cellcolor{yellow!25} Medium & \cellcolor{yellow!25} Medium    \\ \hline
\end{tabular}
\vspace{1pt} 
\caption{Properties of three different transports that can be used to deliver live video.}
\label{tab:comp}
\end{table} 
\begin{itemize}
    \item \textbf{Security:}
    The first metric is security. For this we consider how easy it would be to hijack a given delivery method to inject malicious code into a client. This is based on our current understanding of available attack vectors.
    
    Native multicast is unencrypted, offers no integrity checks and is thus entirely insecure. On the other hand, HTTPS uses TLS, guaranteeing encryption, integrity and server verification. However, it's noteworthy that while TLS 1.1 is deprecated, a significant number of servers still use TLS 1.2. Similarly, as an extension to QUIC, MCQUIC requires TLS1.3. Accordingly, both HTTPS and MCQUIC offer high security, with MCQUIC potentially offering slightly better protection by enforcing TLS 1.3.
    \item \textbf{Privacy:}
    The second metric we consider is privacy, which focuses on understanding the extent to which third parties, whether on or off path, can learn information about a user. 
    
    As native multicast lacks encryption there is no privacy protection at all. HTTPS provides privacy protection by encrypting all application layer data. Utilizing unicast QUIC could potentially offer even higher privacy guarantees, thanks to additional enhancements, e.g. padding frames that can prevent fingerprinting attacks \cite{reed2017identifying}. While MCQUIC is based on QUIC and also offers these mechanisms, as previously mentioned, multicast has inherently weaker privacy. This is due to many clients receiving the same packet. Additionally, the multicast group management mechanisms currently in use lack encryption, allowing on-path nodes to learn which multicast channels a specific subnet is subscribed to. 
    \item \textbf{Deployability:}
    Deployability is the third metric. It estimates how feasible it is to implement and deploy the specific technology to allow for the reception of live video by a user. Given the widespread usage of HTTPS and DASH/HLS, their deployability is well established. We have already discussed the various challenges native multicast faces with regards to deployment. While it may be feasible in certain scenarios, such as IPTV, in general its deployability remains low. When considering the deployability of MCQUIC, two facets need to be considered. Firstly, the ease of integrating it into end-user devices, particularly browsers and browser-based applications. As mentioned, QUIC is already implemented in all major browsers, and with MCQUIC being an extension thereof, there is a clear pathway for deployment. However, MCQUIC's scalability also highly depends on a networks support for multicast. Therefore, its deployability, at least when not having to rely on unicast fallback, is directly influenced by ISPs enabling multicast. The adoption of technologies like BIER by many operators is a promising sign that more and more ISPs might enable multicast in their networks. 
    \item \textbf{Scalability:}
    Finally, we consider how scalable the three methods are. Native multicast imposes no additional per-user overhead whatsoever. Its demand on both the server ingesting the video and the network remains constant regardless of the number of receivers. The HTTPS-based transport incurs high per-user overheads. Each connection must be managed individually, and data must be sent separately for each user. MCQUIC strikes a middle ground between the two. Large packets carrying data are transmitted via multicast, removing the need to send them multiple times for multiple receivers. However, each receiver still has a connection that needs to be managed by the server. Additionally, to guarantee integrity, checksums need to be delivered to clients. As mentioned, this can be done either via unicast or multicast through a Merkle tree. We will further analyze this trade-off in the next section. Figure \ref{fig-subs} shows the bandwidth requirements for the three methods, with both methods of integrity shown for MCQUIC. 
\end{itemize}
\subsection{Evaluation of integrity frame overhead}
The requirement for integrity frames to provide checksums for packets sent over multicast introduces additional overhead. In this section, we will show how big this overhead is and how it can be reduced by sending the frames over the multicast channel utilizing a Merkle tree.

We make the following assumptions for our calculations:
\begin{itemize}
\item The MTU is the commonly used 1500 bytes.
\item QUIC connection ID and packet number are both averaged out to the middle of their respective ranges, i.e., 80 and 16 bits respectively.
\item IPv4 is used, and the header has no additional options, so it is 20 bytes in size.
\end{itemize}
As such, in total, we have a QUIC header size of 13 bytes, which, in addition to the UDP header of 8 bytes and the IPv4 header of 20 bytes, leaves 1459 bytes for frames. 
The header of the frame that carries checksums has a size of 14 bytes, leaving 1445 bytes for checksums. We will consider three different algorithms with varying checksum sizes, as shown in Table \ref{tab:SHA}.

\begin{figure}[h]
  \centering
  \includegraphics[width=\linewidth]{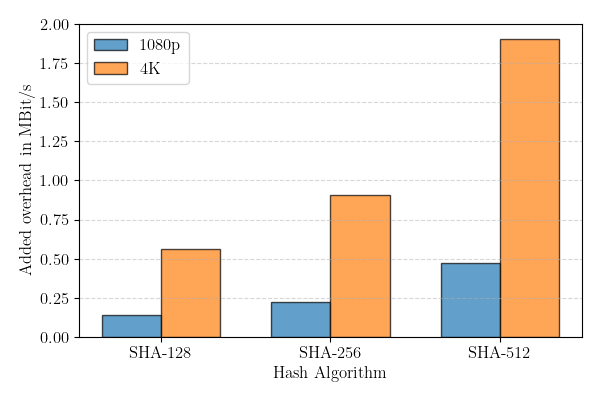}
  \caption{Overhead bandwidth in MBit/s per client for different scenarios.}
   \label{fig-sha}
\end{figure}

\begin{table}[h]
\centering
\begin{tabular}{|l|lll|}
\hline
\rowcolor[HTML]{E0E0E0} 
\textbf{Property} & \textbf{SHA-128} & \textbf{SHA-256} & \textbf{SHA-512} \\ \hline
Checksum size in Bytes         & 20    & 32      & 64      \\ \hline
Number of checksums per packet & 72    & 45      & 22      \\ \hline
Incurred overhead in percent   & 1.4   & 2.3     & 4.8     \\ \hline
\end{tabular}
\vspace{1pt} 
\caption{Comparison of different hash algorithms and their incurred overhead on the connection.}
\label{tab:SHA}
\end{table}
While, as mentioned, SHA-128 is no longer considered resistant to collision attacks, we are including it in our comparison as it might offer an attractive trade-off to further save bandwidth in cases where the probability of collision attacks occurring is low. Applying the calculated overheads to different video streaming scenarios, such as a 1080p and a 4K stream, results in the incurred overheads shown in Figure \ref{fig-sha}, measured in MBit/s per client.
\begin{figure}[h]
  \centering
  \includegraphics[width=\linewidth]{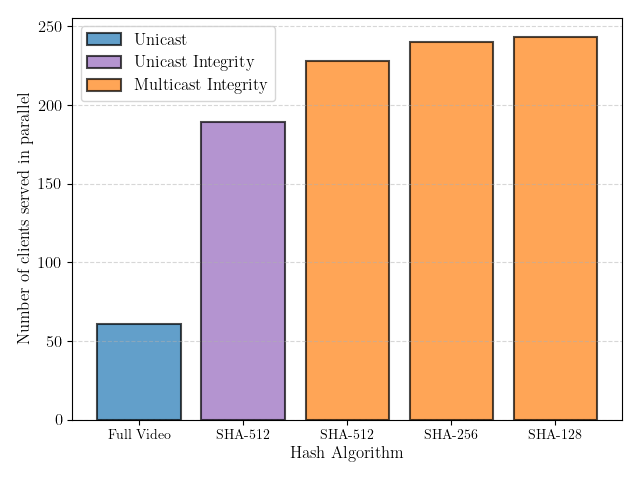}
  \caption{Overhead bandwidth in MBit/s per client for different scenarios.}
   \label{fig-exp}
\end{figure}

If unicast is used to distribute integrity frames, each client will experience the overheads shown. This can still impose significant strain on both network infrastructure and servers. Instead, utilizing the multicast channel to send checksums results in a constant overhead, independent of the number of joined clients. As mentioned earlier, Figure \ref{fig-subs} illustrates the total bandwidth usage in MBit/s for different numbers of subscribers using different delivery methods. As shown, while the impact of including integrity frames is not insignificant on a per-client basis, using multicast to transmit them along with the video segments virtually eliminates it at the scales envisioned for MCQUIC deployments.
\subsection{Case study and experiment}
To compare the benefits of delivering video segments over multicast versus unicast, we conducted an experiment to analyze how many viewers a single server can serve simultaneously. Using aioquic and Python, we implemented a simple QUIC server with two modes. In the first mode, it perpetually sends 1MB-sized video segments to clients over unicast QUIC, while in the second mode, it only computes and sends  the checksums for those segments. The three previously mentioned hashing algorithms were also applied here. For the multicast integrity scenarios, only the checksum of the first segment was sent. The server ran on the smallest available virtual machine instance of a cloud computing provider, with clients connecting to it through the Internet from a domestic network. Figure \ref{fig-exp} illustrates how many clients were able to be served simultaneously. It can be observed that while the improvement is less significant than in theoretical bandwidth savings, there is still a more than fourfold increase in servable clients. Additionally, due to the high amount of overhead incurred in maintaining the QUIC connection itself, the choice of hash algorithm is less relevant. It should be noted that few optimizations were made, and aioquic is not performance-oriented to begin with. Running actual production-level implementations would likely reduce the basic QUIC connection overhead significantly.

\section{Related work}
\label{related}
As mentioned in the introduction, multicast has recently received much interest again. One major ongoing effort is the TreeDN \cite{ietf-mops-treedn-00} project championed by Juniper. It tries to solve a similar problem of delivery of live video content via multicast. However, instead of aiming at a browser based implementation it mainly makes use of AMT and the VLC media player. It provides the ability to stream videos from the MBONE to domestic networks that do not support native multicast reception by going through an AMT relay. It also provides a content portal, similar to YouTube or Twitch, on which content can be found and streamed. Finally, it includes a service that can automatically translate a unicast stream to multicast and ingest it into the MBONE. 

The second major ongoing work relating to multicast is the mentioned BIER \cite{rfc8279}, which aims to reduce state in multicast routing. Instead of the traditional approach of constructing multicast trees, it works by separating the Internet into multiple BIER domains. These could take the form of a company network, an autonomous system or even an entire ISPs network. 
If a multicast packet enters a BIER domain through an ingest router, it gets tagged with IDs for all routers, either inside the domain or on its edge, it has to be delivered to. This means that routers no longer have to keep state for multicast trees as all the necessary information for routing is included in the BIER header of each individual packet. BIER has been implemented and evaluated in P4 \cite{merling2021hardware}.  The downside to this approach is that unlike PIM, it is not a general purpose protocol but instead requires a different mechanism for the tagging of packets at the ingress router depending on the use base, e.g. there are BIER extensions for IS-IS \cite{rfc8401}, OSPF \cite{rfc8444}, multicast VPN \cite{rfc8556} etc.

In general, the advantages of using multicast for content delivery have been well known. Research in this area reaches from multicast in mobile networks \cite{jelger2002multicast} over using overlay networks to connect multicast islands \cite{jin2009island} to multimedia conferencing with multicast~\cite{mccanne1999scalable}.
\section{Conclusion}
\label{conclusion}
In this paper, we introduced a new extension to the QUIC transport protocol that enables the multicast delivery of mass media content. It overcomes many of the obstacles faced by traditional multicast implementations and deployments by leveraging a base unicast QUIC connection as a security anchor for multicast channels. In particular, MCQUIC allows operators to profit from the established efficiency and scalability benefits of multicast, while providing improved security guarantees (e.g., through packet integrity) as well as user privacy.
It does so while introducing only a small overhead, particularly in terms of bandwidth. Additionally, applications can use MCQUIC without many special considerations as it offers automatic unicast fallback.

While there are many future research opportunities in MCQUIC itself, it also allows us to reevaluate other technologies that have so far been deemed unsuitable or inefficient in unicast scenarios. One example of this is scalable video encoding. While its associated overhead makes it unattractive for unicast delivery schemes, it might be a promising way to combine multiple multicast channels into one media stream. This would allow for more fine grained video quality adjustments, thus delivering the best possible QoE to each client.
We are currently in the process of attempting to standardization of MCQUIC at the IETF. The most recent version is available on the datatracker\cite{jholland-quic-multicast-04}.
{\balance
\bibliographystyle{IEEEtran}
\bibliography{base}
}

\end{document}